\documentclass[12pt]{article}
\usepackage{amsmath,amssymb,color}
\usepackage{dsfont}
\usepackage{cite,epsf}
\usepackage{pstricks}
\usepackage{color}

\usepackage{graphicx,array} 
\newcommand{\be}{\begin{equation}}
\newcommand{\ee}{\end{equation}}
\newcommand{\beq}{\begin{equation}}
\newcommand{\eeq}{\end{equation}}
\newcommand{\bea}{\begin{eqnarray}}
\newcommand{\eea}{\end{eqnarray}}

\newcommand{\ba}{\begin{eqnarray}}
\newcommand{\ea}{\end{eqnarray}}

\begin{document}

\begin{titlepage}
\vspace{10pt}
\hfill
{\large\bf HU-EP-12/49}
\vspace{20mm}
\begin{center}

{\Large\bf  Conformal geometry of null hexagons\\
for Wilson loops and scattering amplitudes\\[2mm] 
}

\vspace{45pt}

{\large Harald Dorn 
{\footnote{dorn@physik.hu-berlin.de
 }}, Hagen M\"unkler and Christian Spielvogel}
\\[15mm]
{\it\ Institut f\"ur Physik der
Humboldt-Universit\"at zu Berlin,}\\
{\it Newtonstra{\ss}e 15, D-12489 Berlin, Germany}\\[4mm]

\vspace{20pt}

\end{center}
\vspace{10pt}
\vspace{40pt}

\centerline{{\bf{Abstract}}}
\vspace*{5mm}
\noindent
The cross-ratios do not uniquely fix the class of conformally equivalent
configurations of null polygons. 
In view of applications to Wilson loops and scattering amplitudes we
characterise all conformal classes  of null hexagon configurations
belonging to given points in cross-ratio space. At first this is done for
the ordered set of vertices. Including the edges, we then investigate
the equivalence classes under conformal transformations for null hexagons.
This is done both for the set of null hexagons closed in finite domains 
of Minkowski space as well as for the set including those closed via infinity.
   
\end{titlepage}
\newpage


\section{Introduction}
Wilson loops as  non-local gauge invariant, but path dependent, quantities
play a central role in any gauge field theory. A lot of studies in the past 
four decades were devoted to properties for generic contours, both without and 
with cusps or intersections. Often one is also free to switch to Euclidean space. 
However, in particular due to their relation to the infrared behaviour of QCD
(see e.g.\cite{Korchemsky:1987wg}), also the investigation of properties 
characteristic for Wilson loops in Minkowski space with its indefinite metric
has a long history.

This aspect has gained further momentum by the discovery of the correspondence
between Wilson loops for null-polygonal contours and scattering amplitudes
in\\ ${\cal N}=4$ super Yang-Mills gauge theory, first via the AdS-CFT correspondence
 for strong coupling \cite{alday-malda} and soon afterwards also at weak coupling
\cite{drummond,drummond2}. Null-polygonal contours are very special, insofar as
they have no counterpart in Euclidean space. Furthermore, the dependence on the contour reduces to the dependence on the location of the vertices $\{x_j\}$ of the polygon
and then by Poincar$\acute{\mbox{e}}$ invariance to the dependence on the
related Mandelstam-like variables.
 
Conformal invariance, present on the classical level, is broken due to the presence
of ultraviolet divergences. The behaviour of the dimensional regularised Wilson 
loops for null polygons is somehow controlled by anomalous conformal Ward 
identities \cite{Drummond:2007au}. It results in the so-called BDS structure
plus, starting with hexagons, a remainder function which is conformally invariant.
The BDS structure had been invented before for infrared regularised scattering amplitudes \cite{bds}.

Counting the number of parameters of the conformal group, one finds for a
null hexagon in $\mathbb R^{1,3}$ three remaining continuous parameters
specifying classes of conformally equivalent configurations. The standard
choice are the three cross-ratios
\beq
u_1~=~\frac{x_{13}^2~x_{46}^2}{x_{14}^2~x_{36}^2}~,~~~~u_2~=~\frac{x_{24}^2~x_{15}^2}{x_{25}^2~x_{14}^2}~,~~~~u_3~=~\frac{x_{35}^2~x_{26}^2}{x_{36}^2~x_{25}^2}~,\label{cross-ratios}
\eeq
formed out of the Mandelstam variables
\beq
x^2_{ij}:=(x_i-x_j)^2~.\label{Mandelstam}
\eeq 
Since the anomalous conformal Ward identities encode the consequences of
the invariance of the remainder functions only with respect to infinitesimal 
conformal transformation it remains open, whether the invariance  holds also
for all finite transformations. Some early tests with special examples of 
different
hexagon configurations having the same cross-ratios indicated 
agreement \cite{drummond2,Anastasiou:2009kna}. However, beyond some
doubt concerning the conformal invariance of the remainder function,
there is another reason for asking whether the cross-ratios fix its value.
It is the purely 
geometrical question, whether the three cross-ratios fix
the hexagon configuration uniquely, up to conformal maps. As pointed out in
a footnote of  \cite{Gaiotto:2011dt}, the answer is negative. This fact
becomes more explicit by an example of two hexagon configurations
with the same cross-ratios, but one with and the other without a pair of 
crossing edges \cite{Dorn:2011ec}. Furthermore, there it has also been shown, 
that the two loop remainder takes different values in both cases, but they 
are related by a suitable analytic continuation.

The aim of the present paper is a 
complete characterisation, both of all triples of cross-ratios
as well as of all different classes of conformally equivalent
real configurations in $\mathbb R^{1,3}$, which belong to a given allowed
point in the cross-ratio space. Special conformal transformations have a critical 
light cone which is mapped to infinity. If this cone cuts a certain edge of the 
hexagon, the image of this edge connects the two images of the adjacent vertices
via infinity. Therefore, it is appropriate to proceed in two steps. At first to
talk only about the identification of all the conformal invariants for an
ordered set of six points, whose neighbours are light-like separated. This will be done in section 2. Including in a second step the edges, one has two options.
If one wants to keep all conformal transformations in the game, one has to
enlarge the set of null hexagons closed within Minkowski space by those who
are closed only via infinity. On the other side, it seems to us also 
a legitimate question to ask for conformal classes of null hexagons,
closed exclusively in a finite part of $\mathbb R^{1,3}$. Then one has to
disregard special conformal transformations, whose critical light cone
cuts the hexagon. In two parts of section 3 we explore these two different
points of view.   

In section 4 we add some comments on the location of some special hexagon
configurations in the cross-ratio space. After a conclusion section a few 
technical points are collected in five appendices. 

A last comment in this section concerns the nomenclature we are following
below. The full set of conformal transformations is given by the group $O(2,4)/Z_2$,
 with $Z_2$ representing $\{1,-1\}$. However, if we talk about the conformal
group and related conformal classes, we have in mind the component connected to
the identity, i.e. $SO_+(2,4)/Z_2$. \footnote{The other three components
are generated by including  time reversal $T$ and parity transformation $P$.}     
\section{Conformal classification of 6-tupels with light-like separated neighbours}
Minkowski space $\mathbb{R}^{1,3}$ can be mapped conformally and one to one
to the cone \cite{Dirac:1936fq} 
\beq
W_0^2+W_{0'}^2-W_1^2-W_2^2-W_3^2-W_4^2~=~0\label{cone}
\eeq
in projective $\mathbb R P^5_{2,4}$ (i.e. equivalence classes of points
in $\mathbb R^{2,4}\backslash \{0\}$ with $W\sim V~~~\Leftrightarrow$ \\ $W=\lambda V,~~~~\lambda\in
\mathbb R\backslash \{0\}$).~\footnote{We have added a subscript to  $\mathbb{R} P^ 5$ to emphasise the role of the $(2,4)$ metric in the embedding space.}
The map is ($x\in\mathbb{R}^{1,3}$) 
\bea
W^{\mu}&=&\lambda ~x^{\mu},~~\mu=0,1,2,3~,\nonumber\\
W^{0'}&=&\frac{1}{2}\lambda ~(1-x^{\mu}x_{\mu}),~~~~W^{4}~=~\frac{1}{2}\lambda ~(1+x^{\mu}x_{\mu})\label{map}
\eea
and its inverse
\beq
x^{\mu}~=~\frac{W^{\mu}}{W^{0'}+W^{4}}~.\label{map-inverse}
\eeq
For several uses of this formalism see also e.g.\cite{diraccone}, \cite{mack-luescher}, \cite{Gibbons:1994du}.
Conformal infinity of Minkowski space is mapped to points $\lambda W,~~W^{0'}+W^4=0$.

For two arbitrary points $x_i,~x_j$ one gets
\beq
(x_i-x_j)^2~=~\frac{(W_i-W_j)^2}{(W_i^{0'}+W_i^4)(W_j^{0'}+W_j^4)}~=~\frac{-2~W_iW_j}{(W_i^{0'}+W_i^4)(W_j^{0'}+W_j^4)}~.\label{distance}
\eeq
In particular, light-like separated points in $\mathbb{R}^{1,3}$ correspond to points on the cone \eqref{cone},
which are light-like separated in the sense of $\mathbb R^{2,4}$. 

Remarkably, inserting \eqref{distance} into the cross ratios of type \eqref{cross-ratios}, the denominators of the r.h.s. cancel. Thus the cross-ratios can directly be expressed in terms of the distances of the related points on the cone in $\mathbb R ^{2,4}$ 
\beq
\frac{x^2_{ij}~x^2_{kl}}{x^2_{il}~x^2_{kj}}~=~\frac{(W_iW_j)(W_kW_l)}{(W_iW_l)(W_kW_j)}.\label{cross-W}
\eeq

Each special conformal transformation
\beq
x'^{\mu}=\frac{x^\mu+c^\mu x^2}{1+2cx+c^2x^2}\label{spec}
\eeq 
has a critical light cone in $\mathbb{R}^{1,3}$ with tip at $-\frac{c}{c^2}$.
This whole light cone is mapped to conformal infinity. The point on  $\mathbb R P^5_{2,4}$, corresponding to the image of the tip after applying \eqref{spec}, is 
found with appendix A, eq. \eqref{spec-conf} 
\beq
W^N_{\mbox{\tiny tip}}~=~\lambda~ (0,0,0,0,-1,1)~.\label{tip}
\eeq
Note that in this notation $N$ takes the values $0,1,2,3,0',4$.
The point $W_{\mbox{\tiny tip}}$ is invariant under dilatations, translations
and Lorentz transformations of the original Minkowski space  $\mathbb{R}^{1,3}$.

Let now $x_1$ to $x_6$ denote the vertices of our null hexagon and $W_1$ to $W_6$
their images on $\mathbb R P^5_{2,4}$. Then we apply a special conformal transformation,
whose critical light cone has its tip at $x_6$, i.e. $c=-\frac{ x_6}{x_6^2}$. 
This light cone $\{x_6+n\vert~n^2=0\}$ is mapped to conformal infinity with the  $\mathbb R P^5_{2,4}$ image
\beq
\lambda~\left (n^\mu-\frac{2x_6n}{x_6^2}x_6^\mu,-\frac{(x_6+n)^2}{2},\frac{(x_6+n)^2}{2}\right )~.
\nonumber
\eeq
By including an overall rescaling into the prefactor $\lambda$ 
we arrive at $\lambda(w^\mu,-1,1)$ or $\lambda(w^\mu,0,0)$. The last option corresponds
to the exceptional points on the original light cone, which are also light-like
with respect to the origin. We assume that $x_1$ and $x_5$ are not of exceptional
type\footnote{In each neighbourhood of exceptional points one has non-exceptional 
ones. Hence for our purpose of determining the range of the three cross-ratios this is allowed.}. 

After this we have
\beq
W_6=\lambda _6(0,0,0,0,-1,1),~~W_1=\lambda_1(w^\mu_1,-1,1),~~W_5=\lambda_5(w^\mu_5,-1,1)~.\label{561}
\eeq
Starting from a generic configuration the remaining three points $x_2,~x_3,~x_4$ are then still finite in  $\mathbb{R}^{1,3}$. By a translation we shift $x_3$ into the origin, 
which by \eqref{map} means
\beq
W_3=\lambda_3(0,0,0,0,1,1)~.\label{3}
\eeq
Under translations  in Minkowski space $x\mapsto x+a$ (Appendix A, \eqref{translation}) 
$W=\lambda (w,-1,1) $ goes to $ \lambda (w,-(1+aw),1+aw)$. With the same justification, as given in the last footnote, we assume
the non-exceptional case $1-x_3w_j\neq0,~~j=1,5,$ and find after
rescaling, that this translation does not affect the structure of \eqref{561}. 
Since $x_2$ and $x_4$ are light-like separated from $x_3=0$ the corresponding points on  $\mathbb R P^5_{2,4}$ have
now the form
\beq
W_2=\lambda_2(w_2^\mu,1,1)~,~~~~W_4=\lambda_4(w_4^\mu,1,1)~.\label{24}
\eeq
The cone condition \eqref{cone} and the null condition for the hexagon edges via \eqref{distance} constrain the entries $w^\mu_1,w^\mu_2,w^\mu_4$ and $w^\mu_5$ in \eqref{561} and \eqref{24} by
\beq
w_1w_2~=~w_4w_5~=~2~,~~~~w_j^2~=~0~,~~j=1,2,4,5~.\label{small-w}
\eeq

At this stage  we still have the freedom to use Lorentz transformations and dilatations to further specify the $w_j$.
From Appendix A, \eqref{Lorentz} and \eqref{dilatation} we see that the structure \eqref{561}-\eqref{24} is preserved
and the entries $w_j,~j=1,2,4,5$ are transformed like vectors in  $\mathbb{R}^{1,3}$. 
For generic cases $(x_2-x_4)^2\neq 0$ one has $w_2w_4\neq 0$ and can use a dilatation to achieve $w_2w_4=\pm 2$.
Let us first continue with\\[1mm]
\underline {\it Case A:} $w_2w_4=2$\\[1mm]
Then by a suitable Lorentz transformation we get $w_2=(\pm \vert\vec w\vert,\vec w),~~w_4=(\pm \vert\vec w\vert,-\vec w)$. Rotating the Euclidean unit three vector $\vec w$ in the direction of the 1-axis we
arrive at
\beq
w_2~=~(\pm 1,1,0,0)~,~~~~w_4~=~(\pm 1,-1,0,0)~.\label{small-w24}
\eeq
Now all freedom to map a generic configuration via conformal transformations
to a special subset of configurations has been used. $w_1$ and $w_5$ are
constrained only by \eqref{small-w}, which leads to the structure
\beq
w_1~=~\big (\pm(1+\vec p^{~2}),\vec p^{~2}-1,2\vec p~\big )~,~~~~w_5~=~\big (\pm (1+\vec q^{~2}),1-\vec q^{~2},2\vec q~\big )~.\label{small-w15}
\eeq
Here $\vec p$ and $\vec q$ are arbitrary two-dimensional Euclidean vectors.

Now using \eqref{561}-\eqref{24}, \eqref{small-w24}, \eqref{small-w15} 
and \eqref{cross-W} for the evaluation of the cross-ratios \eqref{cross-ratios}
we get 
\beq 
u_1~=~\frac{1}{1-\vec p^{~2}}~,~~~~u_2~=~\frac{1+\vec p^{~2}\vec q^{~2}-2~\vec p~\vec q}{(1-\vec p^{~2})(1-\vec q^{~2})}~,~~~~u_3~=~\frac{1}{1-\vec q^{~2}}~.\label{u-ears}
\eeq
A similar analysis for\\[1mm]
\underline{\it Case B:} $w_2w_4=-2$\\[1mm]
leads to
\bea
w_2&=&(\pm 1,1,0,0)~,~~~~w_4~=~(\mp 1,1,0,0)~,\nonumber\\
w_1&=&\big (\pm(1+\vec p^{~2}),\vec p^{~2}-1,2\vec p~\big )~,~~~~w_5~=~\big (\mp (1+\vec q^{~2}),\vec q^{~2}-1,2\vec q~\big )~,\label{B-small-w}
\eea
as well as
\beq 
u_1~=~\frac{1}{1+\vec p^{~2}}~,~~~~u_2~=~\frac{1+\vec p^{~2}\vec q^{~2}+2~\vec p~\vec q}{(1+\vec p^{~2})(1+\vec q^{~2})}~,~~~~u_3~=~\frac{1}{1+\vec q^{~2}}~.\label{u-bag}
\eeq

\begin{figure}[h!]
 \centering
 \includegraphics[width=7cm]{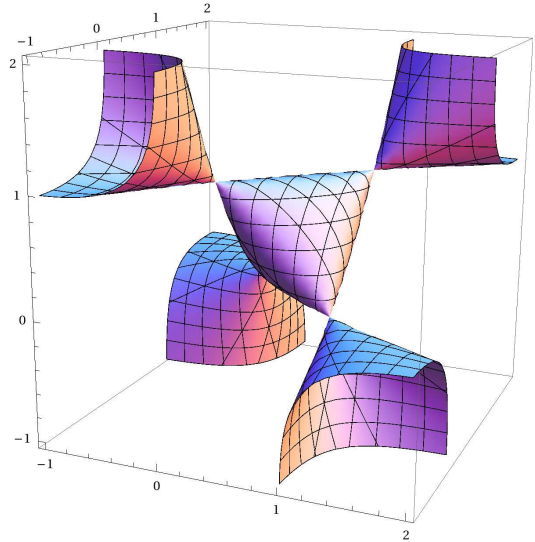} 
\caption{{\it The allowed region for $\mathbb R^{1,3}$ null hexagons in cross-ratio space.}}
 \label{bag-ears}
\end{figure}

The domain in the three-dimensional space of cross-ratios, which can be realised
by real null hexagon configurations in Minkowski space, is fully covered by 
\eqref{u-ears} and \eqref{u-bag} varying independently the three variables
$p,q,z$ ($p=\vert\vec p\vert,~q=\vert\vec q\vert ~, z=\frac{\vec p\vec q}{pq}$ ) within $p,q\geq 0$ and $-1\leq z\leq 1$. Expressing $p$ and $q$ in terms of $u_1$ and $u_3$ one finds ($\mp $ refers to case A/B)
\beq
u_2~=~2 u_1u_3-u_1-u_3+1\mp 2z\sqrt{u_1u_3(1-u_1)(1-u_3)}~.
\eeq 
This domain is depicted in fig.\ref{bag-ears}. 

Its characterisation in terms of the $u_j$ alone
is given by the overall inequality
\beq 
4~u_1u_2u_3-(u_1+u_2+u_3-1)^2~\geq ~0~,\label{domain-ineq}
\eeq
valid in the whole domain. Its five parts, in addition, are specified by 
\bea
\mbox{``bag''}&:&~~~~0\leq u_k\leq 1,~~k=1,2,3~,\label{bag}\\
\mbox{``ear j''},~~j=1,2,3 &:&~~~~u_j\geq 1,~~u_k\leq 0,~~k\neq j~,\label{ear-j}\\
\mbox{``ear 4''}&:&~~~~u_k\geq 1,~~k=1,2,3~.\label{ear-4}
\eea
The central $\it{ bag}$ is realised with case B. 
The four {\it ears} are realised with case A.
 
Note also that the two options, which one has both in case A and B by the choice of the upper/lower signs in \eqref{small-w24}, \eqref{small-w15} and \eqref{B-small-w}, are related by time reversal in Minkowski space, corresponding to $W^0\rightarrow -W^0$ in $\mathbb R^{2,4}$. 

In reconstructing the 2-dimensional vectors $\vec p$ and $\vec q~$  from
the cross-ratios, i.e. from $\vec p^{~2},~\vec q^{~2},~\vec p\vec q$, see \eqref{u-ears},\eqref{u-bag}, besides the freedom of rotations, one has the option
$\vartheta\rightarrow -\vartheta$ for the angle between $\vec p$ and $\vec q$. 
Furthermore, in \eqref{small-w24},\eqref{small-w15} the three 3D vectors $\vec w$, $\vec w_1$ and $\vec w_5$ are involved. After we decided to rotate $\vec w$ in the direction of the 1-axis, the 3D orientation of this triple of vectors manifests itself in
the 2D orientation of the pair $\vec p$, $\vec q$ parameterising $\vec w_1$ and $\vec w_5$. Hence a change of this orientation is due to a parity transformation in Minkowski space.

Finally, one should add the trivial observation, that under a cyclic shift 
of the numbering of the points $\{x_j\}$ one has
\beq
(x_1,x_2,x_3,x_4,x_5,x_6)\mapsto (x_6,x_1,x_2,x_3,x_4,x_5)~~~
\Longrightarrow ~~~(u_1,u_2,u_3)\mapsto (u_3,u_1,u_2)~.\label{shift}
\eeq
Such a shift maps both the bag and ear 4 to itself and implies a corresponding
cyclic mapping of the ears 1 to 3. 

Altogether  we have shown that each ordered set of six light-like separated points 
$(x_1,\dots ,x_6)$ in Minkowski 
space can be mapped by conformal transformations to one of the four standard 
configurations, whose image on  $\mathbb R P^5_{2,4}$ has been described above.
Since case A and B cover different regions in the space of cross-ratios, this
implies that a given generic point $(u_1,u_2,u_3)$ in the allowed region 
\eqref{domain-ineq}-\eqref{ear-4} fixes the position of  $(x_1,\dots ,x_6)$ uniquely, up to conformal transformations, up to the cyclic permutation
$(x_1,x_2,\dots ,x_6)\rightarrow (x_4,x_5,\dots ,x_3)$ and up to time reversal and up to a parity transformation.
\footnote{At non-generic points the degeneracy may be higher, e.g. if all $u_j$ are
equal, each cyclic permutation is included. At the vertex points of the bag
cases A and B are both possible.}
\section{Conformal classification of null hexagons}
\subsection{Null hexagons closed via infinity included }
Topologically the null geodesic, which is determined by a given  pair of 
two consecutive points $x_j,x_{j+1}$ out of a 6-tuple of light-like separated 
points, is a circle on conformally compactified Minkowski space, 
i.e. on $\mathbb R P^5_{2,4}$, see Appendix B. The two points $x_j,x_{j+1}$ divide it 
into two parts. One of them contains a point at conformal infinity. 

Let us first take into consideration also null hexagons which close via infinity,
i.e. allow for the edge connecting $x_j$ and $x_{j+1}$ either the direct connection
or the connection via infinity. Although crossing infinity is not a conformally
invariant characteristic, it is nevertheless possible to mark the two parts
of such a null geodesic in an  $SO_+(2,4)$  invariant way. The notion of a time arrow
on null geodesics is   $SO_+(2,4)$ invariant ($\frac{dx^0}{ds}>0\Rightarrow \frac{d(x')^{0}}{ds}>0$). Therefore,  the two parts can be
distinguished by the alternative of going from  $x_j$ to $x_{j+1}$ with or
opposite to the time arrow.

After this remark we get for each set of cross-ratios
$ 4\times 2^6= 256$ conformally inequivalent classes of hexagon configurations.
Here the first factor  4 corresponds to time reversal and parity and the factor $2^6$
to the choice of one of the two parts of the null geodesics for all 6 pairs of
consecutive vertices of the null hexagon.

Under special conformal transformations 
\beq
(x'-y')^2~=~\frac{(x-y)^2}{(c^2)^2~(x+c/c^2)^2~(y+c/c^2)^2}~.\label{spec-conf2}
\eeq
Therefore, $J_1$ and $J_2$, defined by
\beq
J_1~:=~\mbox{sign}\big (x^2_{13}~x^2_{35}~x^2_{51}\big )~,~~~J_2~:=~\mbox{sign}\big (x^2_{24}~x^2_{46}~x^2_{62}\big )~\label{J}
\eeq 
are conformal invariants. According to our result in the previous section
they cannot contain information independent of the cross-ratios. 

For later
use it is instructive to see this more explicitly. At first from the definition
of the cross-ratios in \eqref{cross-ratios} one gets $J_1~J_2=\mbox{sign}(u_1u_2u_3)$.\\ Eqs.\eqref{bag}-\eqref{ear-4} 
imply 
$\mbox{sign}(u_1u_2u_3)\geq 0$, which means for the generic situation
of all $x^2_{j,j+2}\neq 0$ (Here and below the indices are understood modulo 6 and the index of the cross-ratios modulo 3.)
\beq
\prod _{j=1}^6x^2_{j,j+2}~> 0 ~~~~\mbox{and}~~~~J_1~=~J_2~.\label{JJ}
\eeq
Finally, \eqref{distance} allows to express $J_1$ and $J_2$ in terms of the
$W_j$ (as for the cross-ratios the denominators on the r.h.s. of \eqref{distance} turn out to be  
irrelevant)
\beq
J_1~=~-\mbox{sign}\Big ((W_1W_3)~(W_3W_5)~(W_5W_1)\Big )~,\label{JW}
\eeq
and similarly for $J_2$. Now the $J$'s can be read off directly from the standard
configurations in $\mathbb R P^5_{2,4}$, we find
\bea
J_1~=~J_2&=&-1~,~~~~\mbox{case A}~,\nonumber\\
J_1~=~J_2&=&+1~,~~~~\mbox{case B}~.\label{JAB}
\eea
Since case A and B cover different regions of the cross-ratio space (ears versus
bag), $J_1$ and $J_2$ yield no independent information on conformal equivalence classes (except at the four points where the bag and ears touch each other).
\subsection{Null hexagons closing via infinity excluded} 
In the following we ask a modified question. We are now interested in the set 
of null hexagons which are closed without going via infinity. 
Then we have to restrict the allowed conformal transformations to those
which do not map out of this subset. As a bonus, we now can use the invariance of the 
signs of all the Mandelstam variables 
to characterise different classes of conformally equivalent 
configurations belonging to a given allowed point in the cross-ratio space. 
This sign invariance follows from the observation, that a sign change of one of 
the $x^2_{ij}$ is necessarily connected
with a map to a hexagon configuration passing conformal infinity. This is shown
in appendix E.

Concerning special conformal transformations \eqref{spec-conf2} 
implies that the distance between two not null separated points  
changes iff both points are on different sides of the critical light cone 
centred at $-c/c^2$. This means that special conformal transformations, whose critical light cone cuts any edge of the  to be transformed hexagon, have to be excluded.

As explained in the introduction, via the correspondence between
Wilson loops and gluon scattering amplitudes, our hexagons are also
relevant for 6-point scattering amplitudes with momenta $\{p_j\vert j=1,\dots ,6\}$ related to the edges of the hexagon via 
\beq
p_j:=x_{j+1}-x_j~.\label{xp}
\eeq
Since
\bea
x^2_{jj+2}~>~0~~~~&\Leftrightarrow &~~\mbox{sign}(p_j^0)~=~\mbox{sign}(p_{j+1}^0)~,\nonumber\\
x^2_{jj+2}~<~0~~~~&\Leftrightarrow &~~\mbox{sign}(p_j^0)~=~-\mbox{sign}(p_{j+1}^0)
~,\label{signovernext}
\eea
one can relate the sequence of signs for all the Mandelstam variables 
$x^2_{jj+2}$ to the type of related scattering: $3\rightarrow 3$ or
$2\leftrightarrow 4$ scattering. Then we find by explicit inspection of
all possible cases, that all contributions to $3\rightarrow 3$ yield
$J_1=J_2=-1$ and all contributions to $2\leftrightarrow 4$ scattering
yield $J_1=J_2=+1$. Therefore, we can supplement \eqref{JAB} by
\bea
\mbox{ears 1 to 4}~&\Leftrightarrow ~& \mbox{case A}~~\Leftrightarrow~~3\rightarrow 3~~\mbox{scattering}~,\nonumber\\
\mbox{bag}~&\Leftrightarrow ~& \mbox{case B}~~\Leftrightarrow~~2\leftrightarrow 4~~\mbox{scattering}~.\label{ABscatt}
\eea
Another useful fact is
\beq
x^2_{jj+2}~>~0~,~~~x^2_{j+1j+3}~>~0~~~~\Rightarrow ~~~x^2_{jj+3}~>~0~.\label{jsigns}
\eeq
We now want to list all sign assignments (realisable by real configurations in 
Minkowski space)  to the six Mandelstam variables of type $x^2_{jj+2}$ and the 
three of type $x^2_{jj+3}$. The choice for the first type fixes the correlation
to the type of scattering. Therefore, we collect them together in sets with 
fixed signs for the $x^2_{jj+2}$. Due to the first condition in \eqref{JJ}, for them 
only an even number of negative signs is allowed. In some cases \eqref{jsigns}
fixes already also the sign of some of the $x^2_{jj+3}$.
\footnote{We discuss here the generic case, where all Mandelstam variables
under discussion are different from zero, the degenerated cases will be commented below.}
\beq
\mbox{\bf Sets of sign assignments to the}~ x^2_{jj+2} \label{tab1}
\eeq
\begin{tabular}{|p{10mm}|p{25mm}|p{40mm}|p{25mm}|p{25mm}|}

\hline

 {\bf case} & $x^2_{13}, \dots , x^2_{62}$ & already fixed $x^2_{jj+3}$ & scattering & cycl. perm. \\
\hline
\hline
{\bf a}  &$-~-~-~-~-~-$& $~~~none$ & $~~~3\rightarrow 3$&$~~~~~~~1$\\
\hline
{\bf b}  &$+~+~-~-~-~-$& $~~~x^2_{14}>0$ & $~~~2\leftrightarrow 4$&$~~~~~~~6$\\
\hline
{\bf c}  &$+~-~+~-~-~-$& $~~~none$ & $~~~3\rightarrow 3$&$~~~~~~~6$\\
\hline
{\bf d}  &$+~-~-~+~-~-$& $~~~none$ & $~~~2\leftrightarrow 4$&$~~~~~~~3$\\
\hline
{\bf e}  &$-~+~-~+~+~+$& $~~~x^2_{14},x^2_{52}>0$ & $~~~2\leftrightarrow 4$&$~~~~~~~6$\\
\hline
{\bf f}  &$-~+~+~-~+~+$& $~~~x^2_{52}>0$ & $~~~3\rightarrow 3$&$~~~~~~~3$\\
\hline
\end{tabular}\\[2mm]
We have skipped cases $(++++++)$ forbidden by $\sum_j p_j=0$ and 
$(--++++)$, which would correspond to an unphysical $1\leftrightarrow 5$
transition. As long as we talk about a fixed numbering of the hexagon vertices,
cases b), c) and e) each are one of 6 cyclic permutations of sign assignments
(at fixed numbering of the vertices!).
Cases d) and f) each are one of 3 cyclic permutations, case a) is unique in this respect. These numbers are indicated in the last column of the table.

Now we list for each case of this table the {\it remaining} possibilities
for signs for $x^2_{14},x^2_{25},x^2_{36}$. This fixes then the signs
of the cross-ratios, and we can locate the related position in cross-ratio
space (bag, ears), see \eqref{bag}-\eqref{ear-4}. If one of the cross-ratios is negative, the location is evident. In the case of all cross-ratios positive, the
separation between bag and ear 4 can be made, since we already know,
that the bag is possible only with  $2\leftrightarrow 4$ and ear 4 only
with $3\rightarrow 3$.
\beq
\mbox{\bf case a), sign assignments to the}~ x^2_{jj+3} \label{taba}
\eeq
\begin{center}
\begin{tabular}{|p{20mm}|p{35mm}|p{40mm}|}
\hline
{\bf case} & $~~~~x^2_{14},x^2_{25},x^2_{36}$& location in $u$-space \\
\hline
\hline
{$~~~$\bf a1} & $~~~~~+~+~+$ &$~~~~$ ear 4\\
\hline
{$~~~$\bf a2} & $~~~~~-~-~-$ &$~~~~$ ear 4\\
\hline
$~~~${\bf a3} & $~~~~~-~-~+$ &$~~~~$ ear 2\\
\hline
$~~~${\bf a4} & $~~~~~-~+~-$ &$~~~~$ ear 1\\
\hline
$~~~${\bf a5} & $~~~~~+~-~-$ &$~~~~$ ear 3\\
\hline
$~~~${\bf a6} & $~~~~~+~+~-$ &$~~~~$ ear 2\\
\hline
$~~~${\bf a7} & $~~~~~+~-~+$ &$~~~~$ ear 1\\
\hline
$~~~${\bf a8} & $~~~~~-~+~+$ &$~~~~$ ear 3\\
\hline
\end{tabular}\\[2mm]
\end{center}
For case b) one of the $x^2_{jj+3} $ has already a fixed sign. Three of the
four remaining options would yield cross-ratios located in ears 1 to 3.
Since this pattern has no realisation by real hexagon configurations, we
are left with
\beq
\mbox{\bf case b), sign assignments to the}~ x^2_{jj+3} \label{tabb}
\eeq
\begin{center}
\begin{tabular}{|p{20mm}|p{35mm}|p{40mm}|}
\hline
{\bf case} & $~~~~x^2_{14},x^2_{25},x^2_{36}$& location in $u$-space \\
\hline
\hline
$~~~${\bf b} & $~~~~~+~-~-$ &$~~~~$ bag\\
\hline
\end{tabular} \\[2mm]
\end{center}  
In case c) one finds again 8 possibilities
\beq
\mbox{\bf case c), sign assignments to the}~ x^2_{jj+3} \label{tabc}
\eeq
\begin{center}
\begin{tabular}{|p{20mm}|p{35mm}|p{40mm}|}
\hline
{\bf case} & $~~~~x^2_{14},x^2_{25},x^2_{36}$& location in $u$-space \\
\hline
\hline
{$~~~$\bf c1} & $~~~~~+~+~+$ &$~~~~$ ear 2\\
\hline
{$~~~$\bf c2} & $~~~~~-~-~-$ &$~~~~$ ear 2\\
\hline
$~~~${\bf c3} & $~~~~~-~-~+$ &$~~~~$ ear 4\\
\hline
$~~~${\bf c4} & $~~~~~-~+~-$ &$~~~~$ ear 3\\
\hline
$~~~${\bf c5} & $~~~~~+~-~-$ &$~~~~$ ear 1\\
\hline
$~~~${\bf c6} & $~~~~~+~+~-$ &$~~~~$ ear 4\\
\hline
$~~~${\bf c7} & $~~~~~+~-~+$ &$~~~~$ ear 3\\
\hline
$~~~${\bf c8} & $~~~~~-~+~+$ &$~~~~$ ear 1\\
\hline
\end{tabular}\\[2mm]
\end{center}
In case d) six of the eight options for the $x^2_{jj+3}$ indicated in 
table \eqref{tab1}
would give points in the ears, hence not realisable for a $2\leftrightarrow 4$
configuration.\footnote{The seemingly strange notation in the following table
anticipates the later finding, that case d*) cannot be realised.}
\beq
\mbox{\bf case d), sign assignments to the}~ x^2_{jj+3} \label{tabd}
\eeq
\begin{center}
\begin{tabular}{|p{20mm}|p{35mm}|p{40mm}|}
\hline
{\bf case} & $~~~~x^2_{14},x^2_{25},x^2_{36}$& location in $u$-space \\
\hline
\hline
{$~~~$\bf d*} & $~~~~~+~+~+$ &$~~~~$ bag \\
\hline
{$~~~$\bf d} & $~~~~~-~-~-$ &$~~~~$ bag\\
\hline
\end{tabular}\\[2mm]
\end{center}
Case e) appears in table \eqref{tab1} with two already fixed signs for
the $x^2_{jj+3}$. Only one of the two remaining option gives points in the bag.
\beq
\mbox{\bf case e), sign assignments to the}~ x^2_{jj+3} \label{tabe}
\eeq
\begin{center}
\begin{tabular}{|p{20mm}|p{35mm}|p{40mm}|}
\hline
{\bf case} & $~~~~x^2_{14},x^2_{25},x^2_{36}$& location in $u$-space \\
\hline
\hline
{$~~~$\bf e} & $~~~~~+~+~-$ &$~~~~$ bag \\
\hline
\end{tabular}\\[2mm]
\end{center} 
Finally, for case f) table \eqref{tab1} allows four options.
\beq
\mbox{\bf case f), sign assignments to the}~ x^2_{jj+3} \label{tabf}
\eeq
\begin{center}
\begin{tabular}{|p{20mm}|p{35mm}|p{40mm}|}
\hline
{\bf case} & $~~~~x^2_{14},x^2_{25},x^2_{36}$& location in $u$-space \\
\hline
\hline
{$~~~$\bf f1} & $~~~~~+~+~+$ &$~~~~$ ear 4 \\
\hline
{$~~~$\bf f2} & $~~~~~-~+~-$ &$~~~~$ ear 1 \\
\hline
{$~~~$\bf f3} & $~~~~~-~+~+$ &$~~~~$ ear 3 \\
\hline
{$~~~$\bf f4} & $~~~~~+~+~-$ &$~~~~$ ear 2 \\
\hline
\hline
\end{tabular}\\[2mm]
\end{center}

So far we have used our previous results \eqref{ABscatt} and \eqref{jsigns} to
eliminate some sign options. It remains to decide, whether there exist further 
restrictions or whether all options listed in tables  \eqref{taba}-\eqref{tabf}
indeed can be realised by real configurations in Minkowski space. If a certain sign
option is allowed, we also want to know for sure, whether this then
holds for all points in cross-ratio space, either in the ears ($3\rightarrow 3$) 
or in the bag ($2\leftrightarrow 4$).

By explicit inspection one finds, that all listed sign options, except case a1)
and case d*) correspond to null hexagons, which have at least one vertex $x_j$,
such that the remaining three non-adjacent vertices $x_{j+2},x_{j-2}$ and $x_{j+3}$
are all inside the light cone of $x_j$ or all outside this light cone. This is a 
necessary condition for treating them with standard configurations, which are deformations of certain two-dimensional set-ups as introduced in \cite{bubble}. By scanning
then all options for these standard configurations it turns out, that indeed
all cases beyond the two exceptions can  be covered, for details see appendix C.

We show in appendix D, that the exceptional option of case d*) cannot be realised
by real configurations. However, the other exceptional case a1) is 
possible. An explicit example is presented in appendix D.  

Now we can count the number of inequivalent conformal classes 
for null hexagons (with numbered vertices)  belonging to
a given point in  cross-ratio space. For points in the ears it is 
$1\times 2+6\times2+3\times 1=17$ and for points in the bag $6\times 1
 +3\times 1 +6\times 1=15$. In both cases there is an additional factor
4 due to the still open possibility of overall time reversal  and parity transformation.
\section{Special configurations}
For a hexagon in a two-dimensional $\mathbb R^{1,1}$ subspace one finds 
\beq
\mbox{2D:}~~~(u_1,u_2,u_3)~=~(1,1,1)~.
\label{2D} 
\eeq
If it lives in a three-dimensional $\mathbb R^{1,2}$ subspace, one has $z=\pm1$
instead of $-1\leq z\leq 1$ in the discussion around \eqref{domain-ineq}.
This means that the corresponding point in cross-ratio space has to be on the
surface separating the allowed and not allowed region in fig.\ref{bag-ears}, i.e.
\beq
\mbox{3D:}~~~4~u_1u_2u_3-(u_1+u_2+u_3-1)^2~=~0\label{3D}~.
\eeq

For the degeneration to a pentagon a certain vertex has to coincide
with one of its next neighbours. For generic location of the remaining
vertices \eqref{cross-ratios} leads to 
\beq
\mbox{pentagons:}~~~(u_1,u_2,u_3)~=~(1,0,0)~\mbox{or}~(0,1,0)~\mbox{or}~(0,0,1)~.
\label{pentagon} 
\eeq
The existence of these three pentagon points in \eqref{pentagon} is not a source for
different conformal classes for pentagons. They only reflect the 
freedom one has in choosing the location
for the sixth edge, if one wants to extend a pentagon back to a hexagon.

In a collinear limit one of the $x^2_{jj+2}$ is zero (without having a pentagon
case). Then necessarily one of the cross-ratios is zero, and in addition  \eqref{domain-ineq} implies that the sum of the other two  cross-ratios is equal to one, i.e.
\beq
\mbox{collinear limits:}~~~u_j~=~0 ~\mbox{and}~~ u_{j+1}+u_{j+2}~=~1~,~~~\mbox{for}~j=1,2~\mbox{or}~3~.\label{collinear}\eeq

For hexagons with two crossing edges it has been shown in \cite{georgiou},\cite{Dorn:2011gf },\cite{Dorn:2011ec}, that
\beq
\mbox{crossing edges:}~~~u_j~=~1 ~\mbox{and}~~ u_{j+1}=u_{j+2}~,~~~\mbox{for}~j=1,2~\mbox{or}~3~.\label{crossing}
\eeq
At this point an amusing side remark is in order. As mentioned in the introduction, in an appendix of
\cite{Dorn:2011ec} an explicit example has been discussed for
two hexagon configurations with and without self-crossing, having the same 
cross-ratios of type \eqref{crossing}. It has been
used as an argument in favour of the existence of different conformal
classes for a given set of cross-ratios. From our discussion in section 2
we now know that there must exist a conformal transformation, which
maps the two sets of vertex points to each other. Nevertheless the two
hexagons of that example are not conformally equivalent. They use as edges different
parts of the null geodesics defined by two adjacent vertices. In the case
with no crossing in finite Minkowski space the crossing is just at infinity, 
in the sense discussed in our appendix B.
  
Other interesting limiting cases in cross-ratio space are connected with
multi Regge limits \cite{Lipatov:2010ad,Bartels:2010tx}. They correspond
to a special parametrisation for an approach to the pentagon points \eqref{pentagon},
either from inside the bag ($2\leftrightarrow 4$ scattering) or from ears 1 to 3 ($3\rightarrow 3$ scattering).

\begin{figure}[h!]
 \centering
 \includegraphics[width=12cm]{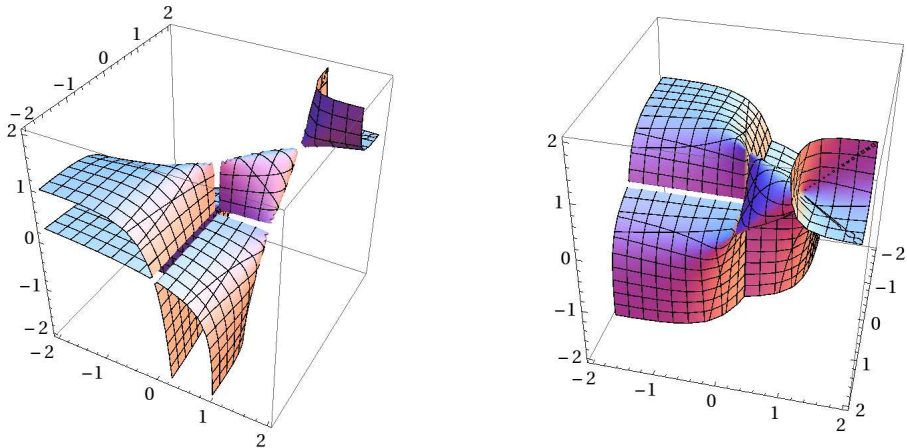} 
\caption{{\it Two views of the allowed region for null hexagons in the $v=1/u$
space. There is still one more component of the bounding surface not shown
in these pictures. It crosses the diagonal at $v_1=v_2=v_3=4$.}} 

 \label{inv}
\end{figure}

Before closing this section we comment on motions in cross-ratio space (see 
figs.\ref{bag-ears} and \ref{inv})
generated by continuous deformations of hexagons. 
Transitions between the bag and the ears 
and among the ears are possible only via sign changes of Mandelstam variables.
The signs of the $x^2_{jj+2}$ are responsible for the distinction between bag and ears.
Besides realisation in passing through the 2D point $(1,1,1)$, a sign change
of one of the $x^2_{jj+2}$ can take place only if one crosses one of the pentagon points
\footnote{They can become zero also by approaching the collinear lines
\eqref{collinear}, but away from the pentagon points there one is faced with
a local maximum/minimum situation.}.

The distinction between the ears is due to the signs of the $x^2_{jj+3}$. If one
of them changes sign by passing zero, two cross ratios change their sign by
passing infinity. To make these transitions more easy on the eyes we depict
in fig.\ref{inv} the allowed cross-ratios in terms of $v_j:=\frac{1}{u_j}~,j=1,2,3$. Ear 4 of fig.\ref{bag-ears} is mapped to the central region in $v$-space. Ear 1,2 and 3 appear again as ears extending into the regions with two
negative $v$ and one positive $v$. The bag of  fig.\ref{bag-ears} is mapped
into part of the region where all $v_j>1$. 

Now the transitions take place on the pieces of the $v_1,~v_2$ or $v_3$ axis
where the images of two ears touch each other.
  
\section{Conclusions}
We have given a complete characterisation of the region in cross-ratio space,
which corresponds to real null hexagon configurations in $\mathbb R^{1,3}$.
To each generic point in the allowed region corresponds, up to time reversal, parity 
and up to the cyclic permutation $(x_1,\dots ,x_6)\rightarrow (x_4,\dots ,x_3)$,
just one class of conformally equivalent ordered sets of six points with null separated neighbours.

Using these six points as vertices of null hexagons one has the option
to include or exclude hexagons which close via infinity. In the first
case one gets then for each class of the ordered 6-tuples $ 4\times 2^6$ conformal
classes of null hexagons. 

Taking into account only null hexagons closed in finite domains of  
$\mathbb R^{1,3}$, one has to exclude special conformal transformations
whose critical light cone crosses the hexagons. Then the signs of the 
Mandelstam variables become invariants. As a result we found 17 
conformal classes for generic points in the cross-ratio region corresponding
to $3\rightarrow 3$ scattering and 15 classes in the region corresponding
to $2\leftrightarrow 4$ scattering (by time reversal  and parity there is still 
a factor 4). 

After our purely geometrical analysis further studies should yield insight
into the following issues. Are the remainder functions of Wilson loops 
for different conformal classes, but the same point in cross ratio space,
related by analytic continuation? A special example for such a continuation
has been analysed in \cite{Dorn:2011ec}, but it has to be clarified whether
analytic continuations connect all classes. Is there some new information
in Wilson loops for contours closed only via infinity? 

For the investigation
of these questions one should also keep in mind, that there exists an alternative
conformal compactification of Minkowski space. It differs from that used in section 2
by restricting oneself in the definition of equivalence classes to $\lambda >0$
\footnote{For an early parallel discussion of both options see \cite{Castell:1969ck}.}. As a consequence, $\mathbb R^{1,3}$ is mapped to half of the modified projective
cone. The version used above is obtained back by identification of antipodal points.
  
After extension to the universal covering, this setting turned out to be appropriate 
for a causal quantisation of
conformally invariant field theories and its relation to the corresponding
Euclidean theory, see \cite{mack-luescher} and refs. therein. 

Our option to allow hexagon configurations closing via infinity comes then
back as the option to play with edges connecting a certain vertex to the antipode
of its neighbour.

Concerning Wilson loops for contours closed only via infinity 
another issue seems to be of interest for further investigation. It is the relation
to the anomaly \footnote{This anomaly has
to be distinguished from the anomaly with respect to infinitesimal
conformal transformations \cite{Drummond:2007au}, which via differential equations governs
for instance the UV divergencies due to cusps
of the contour and also parts of the renormalised Wilson loops.} with respect to inversions studied for Euclidean  ${\cal N}=4$ super Yang-Mills theory in refs. \cite{Drukker,Zarembo}. The Wilson loop for a straight
line is equal to one, but that for a circle equal to a nontrivial function
of the coupling constant, although a circle touching the origin is mapped to a 
straight line
under inversion at the unit sphere. Since in the Euclidean version only one
point of a closed contour passing the origin is mapped to infinity, it has been
argued by locality that the relative factor between Wilson loops before and after the inversion is universal for smooth contours \cite{Drukker}. 

In Minkowski space at least perturbative calculations  should be
feasible before and after application
of a special conformal transformation which opens a closed contour. At the strong coupling end one should be able to handle the situation for instance for null tetragons
where the corresponding AdS string surface is known explicitely \cite{alday-malda}.  

The conformal geometry of null polygons as analysed in the present paper could be
also usefully beyond  ${\cal N}=4$ super Yang-Mills theory, for instance in QCD,
after one succeeds in separating conformal invariant pieces of the corresponding
Wilson loops.
\\[10mm]
\noindent
{\bf Acknowledgement}\\[2mm]
H.D.  thanks G. Jorjadze and  S. Wuttke for helpful discussions and E. Radatz 
for collaboration in an early stage.  
The work is supported in part by DFG via SFB 647 and by 
VolkswagenStiftung via grant I/84600. 
\newpage
\section*{Appendix A}
We are interested in the explicit form of the $SO(2,4)$ transformations acting in $\mathbb R ^{2,4}$,
which correspond to finite translations, Lorentz transformations, dilatations and
special conformal transformations. For some related discussion on the level
of infinitesimal transformations see e.g. \cite{osborn}. If we map from
Minkowski space to $\mathbb R P^5_{2,4}$ with \eqref{map}, perform  $\Lambda\in SO(2,4)$ in $\mathbb R^{2,4}$ and then go back to Minkowski via \eqref{map-inverse} we get 
\beq
x'^{\mu}=\frac{\Lambda^{\mu}_{~\nu}x^{\nu}+\frac{1}{2}(\Lambda^{\mu}_{~0'}+\Lambda^{\mu}_{~4})-\frac{1}{2}(\Lambda^{\mu}_{~0'}-\Lambda^{\mu}_{~4})x^2}{ (\Lambda^{0'}_{~\nu}+\Lambda^{4}_{~\nu})x^{\nu}+\frac{1}{2}(\Lambda^{0'}_{~0'}+\Lambda^4_{~0'}+
\Lambda ^{0'}_{~4}+\Lambda^4_{~4})-\frac{1}{2}(\Lambda ^{0'}_{~0'}+\Lambda^4_{~0'}-\Lambda^{0'}_{~4}-\Lambda^4_{~4})x^2}\label{conftrafo}~.
\eeq
To realise a translation $x'^{\mu}=x^{\mu}+a^{\mu}$ one has to choose $\Lambda^{\mu}_{~\nu}=\delta^\mu_\nu,~~(\Lambda^{\mu}_{~0'}+\Lambda^{\mu}_{~4})=2 a^\mu,~~(\Lambda^{0'}_{~0'}+\Lambda^4_{~0'}+\Lambda ^{0'}_{~4}+\Lambda^4_{~4})=2 $ and the remaining coefficients of $x^\nu$ or $x^2$ in \eqref{conftrafo} equal to zero.
Together with the $SO(2,4)$ condition $\eta^{KL}=\Lambda^K_{~A}\eta^{AB}\Lambda^{L}_{~B}$ this fixes $\Lambda$ up to the freedom $\Lambda\rightarrow -\Lambda$ \footnote{For the capital indices $M,N$ etc. we take
for convenience the ordering $0,1,2,3,0',4$.}
\beq
\mbox{\underline{translation}:}~~\Lambda ^M_{~N}~=~\left (
\begin{array}{ccc}
\mathds{1}&a^\mu&a^\mu\\
-a_\nu&1-\frac{a^2}{2} &-\frac{a^2}{2}\\
a_\nu&\frac{a^2}{2}&1+\frac{a^2}{2}
\end{array}\right )~.\label{translation}
\eeq
In a similar manner we get $x'^{\mu}=\frac{x^\mu+c^\mu x^2}{1+2cx+c^2x^2}$ via
\beq
\mbox{\underline{special conformal}:}~~\Lambda ^M_{~N}~=~\left (
\begin{array}{ccc}
\mathds{1}&-c^\mu&c^\mu\\
c_\nu&1-\frac{c^2}{2} &\frac{c^2}{2}\\
c_\nu&-\frac{c^2}{2}&1+\frac{c^2}{2}
\end{array}\right )~.\label{spec-conf}
\eeq
Lorentz transformations in Minkowski space correspond to
\beq
\mbox{\underline{Lorentz}:}
~~\Lambda ^M_{~N}~=~\left (
\begin{array}{ccc}
\Lambda^\mu_{~\nu}&0&0\\
0&1&0\\
0&0&1
\end{array}\right )~, ~~~~~~~\Lambda^\mu_{~\nu}~\in ~ SO(1,3).\label{Lorentz}
\eeq
And finally, dilatations $x'^\mu=e^{-\rho}x^{\mu}$ have as partner
\beq
\mbox{\underline{dilatation}:}~~\Lambda ^M_{~N}~=~\left (
\begin{array}{ccc}
\mathds{1}&0&0\\
0&\cosh\rho&\sinh\rho\\
0&\sinh\rho&\cosh\rho
\end{array}\right )~.\label{dilatation}
\eeq
\section*{Appendix B}
Here we collect some properties of the images of Minkowski space null geodesics
in  $\mathbb R P^5_{2,4}$. Null geodesic have the form $x+t~n$ with $  x,n\in \mathbb R^{1,3},~n^2=0,~n^0=\pm 1,~~ t\in \mathbb R$. Via \eqref{map} the corresponding image is given by
$W(t)=\lambda (n+\frac{x}{t},\frac{1-x^2}{2t}-xn,\frac{1+x^2}{2t}+xn)$. For $t\rightarrow \pm \infty$ we get the same point, it is
\bea
W(\infty)&=&\lambda~(\frac{n^\mu}{xn},-1,1)~~~ \mbox{for}~ xn\neq 0~,\label{W-inf}\\
W(\infty)&=&\lambda~(n^\mu,0,0)~~~\mbox{for}~ xn= 0~.\nonumber
\eea
Thus null geodesics on conformally compactified Minkowski space
\footnote{Its topology is $S^1\times S^3$, see e.g.\cite{Gibbons:1994du}.}
have the topology of a circle. For two null geodesics, crossing at $x$ with
different directions $n_1,n_2$, the corresponding points at infinity
are different. Two different null geodesics do either not intersect
or intersect at {\it only} one point. 

If one asks for the condition on $x_1,n_1$ and $x_2,n_2$
for crossing at infinity one gets
\beq
n_2^\mu ~=\pm n_1^\mu~,~~~~x_1n_1~=~x_2n_1~.\label{infcros}
\eeq
\section*{Appendix C} 
In \cite{bubble} conformal transformations where used to map a certain
class to the standard configuration $(\Lambda \rightarrow +\infty$)
\bea
x_3=\big (0,0,0\big )~,~x_2=\big (-\frac{1+\vec p^{~2}}{2}-\frac{1-\vec p^{~2}}{2
},\vec p ~\big )~,~x_4=\big (-\frac{1+\vec q^{~2}}{2},\frac{1-\vec q^{~2}}{2},\vec q
~\big )~,\label{bubble}\\
x_1=x_2~+~\big (\Lambda /2,-\Lambda /2,0\big )~,~x_5=x_4~+~\big (\Lambda /2,\Lambda /2,0\big )~,~x_6=\big (\Lambda ,0,0)+\dots ~.
\nonumber
\eea
Note the different role of the two-dimensional transversal vectors $\vec p$ and
$\vec q$ compared to our treatment in section 2. Here they are used to move
the points staying finite into the transversal directions, there they
move transversally the points, which are  put into conformal infinity.

\begin{figure}[h!]
 \centering
 \includegraphics[width=14cm]{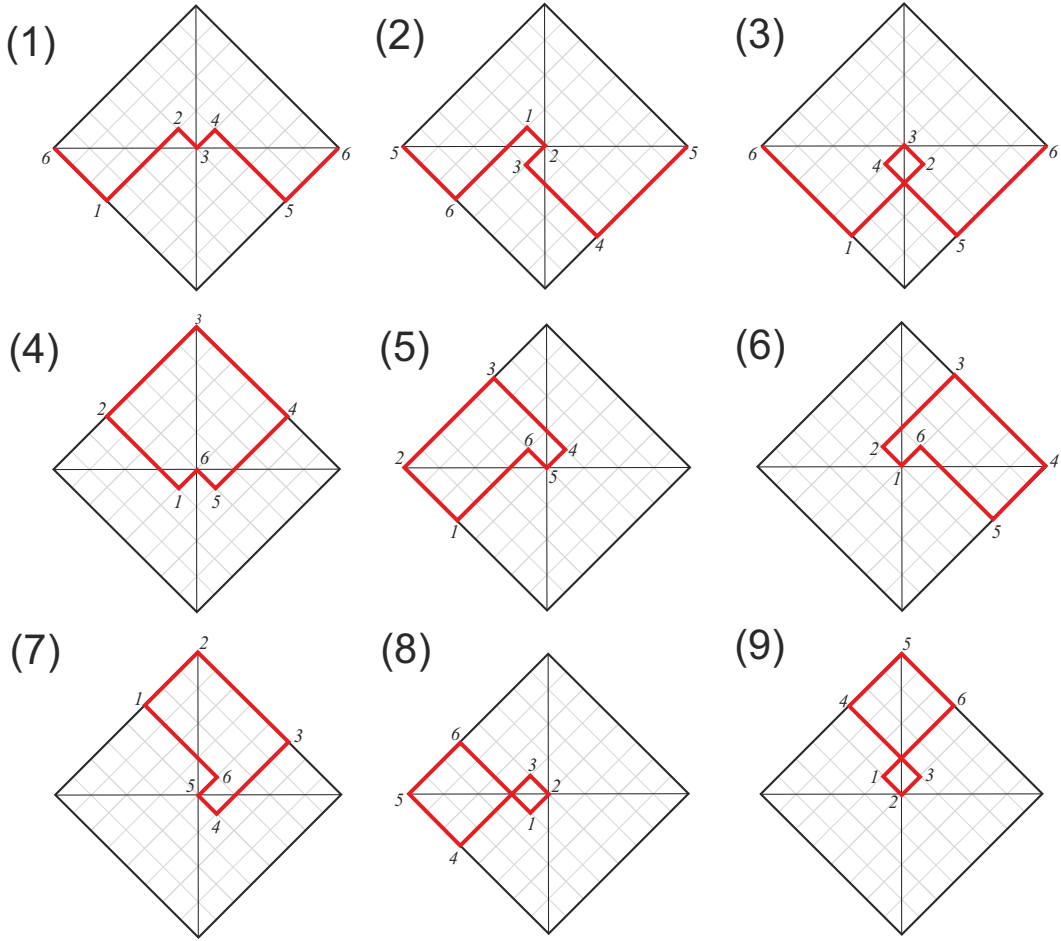} 
\caption{{\it Penrose diagrams of $2D$ projection of standard configurations
at $\vec p=\vec q=0$. The case in} \cite{bubble} {\it and presented in eq. \eqref{bubble}
corresponds to picture} (4) {\it after a suitable cyclic permutation of the vertices.}} 
\label{standards}
\end{figure}

The resulting formulas for the cross-ratios have the same form as \eqref{u-ears}.\\
In the following we need 8 more configurations of a similar nature as that in \eqref{bubble}. We call them 
all standard configurations and show the Penrose diagrams of their projection
on $\mathbb R^{1,1}$ (for $\vec p=\vec q=0$) in fig.\ref{standards}.
The assignments of the vertex
number to the vertices in fig.\ref{standards} is fixed by the sequence
of plus/minus in table \eqref{tab1}. Only in case a) (corresponding to standard configuration 1)) we can make use of cyclically permuted vertex assignments. 

It is straightforward to check that standard configurations 1,3,4,5,6 and 9 yield
cross-ratio formulas looking, up to a possible permutation of the $u_j$, 
as \eqref{u-ears} and standard configurations 2,7 and 8 yield the form \eqref{u-bag}.
Varying $\vec p$ and $\vec q$, each of these 9 standard configurations covers either the complete ear or bag region in 
cross-ratio space. To check which of the classes of tables \eqref{taba}-\eqref{tabf} are realised, we have to consider the sign pattern for the Mandelstam
variables $\{x^2_{jj+2}\}$ and $\{x^2_{jj+3}\}$ in these standard configurations.
In each of these cases the sign of all $\{x^2_{jj+2}\}$ is fixed. Beyond this also the sign of that Mandelstam variable out of 
$\{x^2_{jj+3}\}$, which is related to the point in the origin, is fixed and can be read off from fig.\ref{standards}. The sign of the other two is also fixed
for the bag configurations, but can take independently the values $\pm$
for ear configurations. This yields the following table
\begin{center}
\begin{tabular}{|p{30mm}|p{30mm}||p{30mm}|p{30mm}|}
\hline
{\bf stand. config.} & {\bf conf. classes}&{\bf stand. config.} & {\bf conf. classes} \\
\hline
\hline
{$~~~$\bf 1,} $x_3$ in origin & a2, a4, a5, a6&$~~~~~~~~~${\bf 5} & c2, c3, c5, c7  \\
\hline
{$~~~$\bf 1,} $x_2$ in origin & a2, a3, a5, a7&$~~~~~~~~~${\bf 6} & c2, c3, c4, c8  \\
\hline
{$~~~$\bf 1,} $x_1$ in origin & a2, a3, a4, a8 &$~~~~~~~~~${\bf 7} & e\\ 
\hline
{$~~~~~~~~~$\bf 2} & b& $~~~~~~~~~${\bf 8} & d\\
\hline
$~~~~~~~~~${\bf 3} & c2, c4, c5, c6&$~~~~~~~~~${\bf 9} & f1, f2, f3, f4  \\
\hline
$~~~~~~~~~${\bf 4} & c1, c3, c7, c8& &  \\
\hline
\end{tabular}\\[2mm]
\end{center}

Now we see that all conformal classes of \eqref{taba}-\eqref{tabf}, except
{\bf a1) and d*)}, can be realised by standard configurations sketched in 
fig.\ref{standards}.

\section*{Appendix D} 
We first prove that generically there do not exist real configurations 
of type  d*).
In general, a generic null hexagon configuration can be characterised
in the following way. Consider a hexagon in $\mathbb R^3$, spanned by the
edges $\vec p_1,\dots \vec p_6$. With  $p_j^0=\pm\vert\vec p_j\vert$ one
can see it as the space projection of a cusped null line in $\mathbb R^{1,3}$.
This null line becomes a (closed) hexagon iff the sums of the three-dimensional
lengths of edges, corresponding to the plus/minus choice for the time component,
are equal to each other. 

Applied to the type d*) this imposes the condition 
\beq
\vert\vec p_3\vert +\vert\vec p_6\vert ~=~\vert\vec p_1\vert +\vert\vec p_2\vert+
\vert\vec p_4\vert +\vert\vec p_5\vert~.\label{nod*}
\eeq
To realise the  d*) set-up, all the $x^2_{jj+3}$ would have to be positive.
This would imply, that in a suitable Lorentz frame the space components
of e.g. vertices 1 and 4 coincide: $\vec x_1=\vec x_4$, i.e.
\beq
\vec p_1+\vec p_2+\vec p_3~=0~~~~~\mbox{and}~~~~~\vec p_4+\vec p_5+\vec p_6~=~0~.
\label{nod*1}
\eeq
By the triangle inequality one has $\vert\vec p_3\vert\leq \vert\vec p_1\vert +\vert\vec p_2\vert$ and $\vert\vec p_6\vert\leq \vert\vec p_4\vert +\vert\vec p_5\vert$.
This is in conflict with \eqref{nod*} (of course except in degenerated cases
where both triangles behind \eqref{nod*1} collapse to a line).\\

\begin{figure}[h!]
 \centering
 \includegraphics[width=6cm]{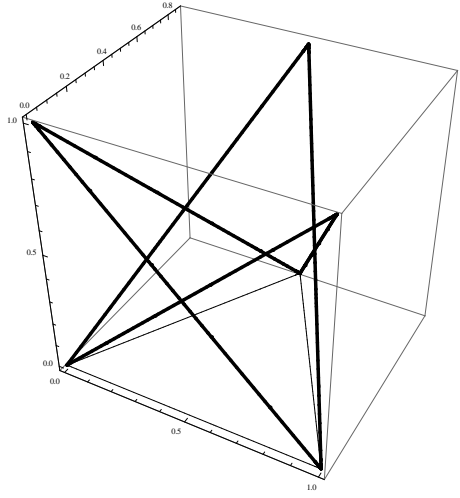} 
\caption{{\it Example for a configuration in class a1). The timelike
coordinate points upwards. With thin lines the projection to the spacelike $(x^1,x^2)$-plane is indicated.} }
\label{special-conf}
\end{figure}

The just used three-dimensional point of view is also helpful to find
explicit examples for real configurations of type a1). Since also here
$x^2_{14}$ is positive, in a suitable Lorentz frame the 3-dimensional
hexagon becomes a pair of triangles, but now, instead of \eqref{nod*1}, with the
condition 
\beq
\vert\vec p_1\vert +\vert\vec p_3\vert+\vert\vec p_5\vert ~=~ \vert\vec p_2\vert +\vert\vec p_4\vert+\vert\vec p_6\vert~.\label{a1ex}
\eeq 
This imposes no obstruction by the triangle equations, and just putting the two 
triangles on top of each other gives an explicit example for a type a1) 
configuration in a $\mathbb R^{1,2}$ subspace
\bea
x_1&=&\big (0,0,0\big )~,~~~x_2~=~\big (1,1,0\big )~,~~~x_3~= ~\big (0,\frac{1}{2}~,\frac{\sqrt{3}}{2}\big )~,\nonumber\\
x_4&=&\big (1,0,0\big )~,~~~x_5~=~\big (0,1,0\big )~,~~~x_6~=~\big (1,\frac{1}{2},\frac{\sqrt{3}}{2}\big )~.\label{examplea1}
\eea
It is depicted in fig.\ref{special-conf}. The related cross-ratios are $u_1=u_2=u_3=1$. 
By rotating the two triangles out of the coincidence position, rotating one of 
them into the third space direction or playing with the lengths of the triangle 
sides, one for sure has enough freedom to reach all points in ear 4 by deformations of the special configuration \eqref{examplea1}.
\newpage
\section*{Appendix E } 
Here we show that, as soon as the sign of $(x_i-x_j)^2$ changes under a conformal 
transformation, the corresponding image of each continuous contour connecting 
$x_i$ and $x_j$ passes infinity. Let us denote by $W_i$ and $W_j$ the related points on 
the cone in $\mathbb R^{2,4}$. Then $\mbox{sign}\big ((x_1'-x_2')^2\big )=-\mbox{sign}\big ((x_1-x_2)^2\big )$ via \eqref{distance} implies ($\Lambda\in SO(2,4)$)
\bea
\mbox{sign}\big ((\Lambda W_1)^{0'}+(\Lambda W_1)^{4}\big )&=&\pm ~\mbox{sign}\big (
W_1^{0'}+W_1^4\big )~,\nonumber\\
\mbox{sign}\big ((\Lambda W_2)^{0'}+(\Lambda W_2)^{4}\big )&=&\mp ~\mbox{sign}\big (
W_2^{0'}+W_2^4\big )~.\label{a}
\eea
Let us now consider the function 
\beq
f_{\Lambda}(W):=\frac{(\Lambda W)^{0'}+(\Lambda W)^4}{W^{0'}+W^4}~
\eeq
and evaluate it along a continuous contour from $W_1$ to $W_2$, whose image
in $\mathbb R^{1,3}$ stays in a finite region. Due to \eqref{a} it has opposite sign on 
both ends. Hence one necessarily finds zeros along the contour. These zeros of $(\Lambda W)^{0'}+(\Lambda W)^4$ correspond 
to passing conformal infinity of $\mathbb R^{1,3}$ . 

\newpage 
 
\end{document}